\documentclass[onecolumn,showpacs,preprintnumbers,amsmath,amssymb,floatfix]{revtex4}

\usepackage{graphicx}
\usepackage{dcolumn}
\usepackage{bm}

\newcommand{\beq}{\begin{equation}}
\newcommand{\eeq}{\end{equation}}
\newcommand{\bey}{\begin{eqnarray}}
\newcommand{\eey}{\end{eqnarray}}

\begin{document}

\title{Viability of Bianchi type V universe in
$f(R,T)= f_{1}(R)+f_{2}(R)f_{3}(T)$ gravity}

\author{Lokesh Kumar Sharma}
\email{lksharma177@gmail.com} \affiliation{Department of Physics, GLA University, Mathura - 281406, India}
\author{Benoy Kumar Singh}
\email{benoy.singh@gla.ac.in} \affiliation{Department of Physics, GLA University, Mathura - 281406, India}
\author{Anil Kumar Yadav}
\email{abanilyadav@yahoo.co.in} \affiliation{Department of
Physics\\ United College of Engineering and Research, Greater Noida - 201310, India}


\begin{abstract}
In this paper, we examine the viability of Bianchi type V universe in $f(R,T)$ theory of gravitation.
To solve the field equations, we have considered the power law for scale factor and constructed a singular Lagrangian model which is based on the coupling between Ricci scalar R and trace of energy-momentum tensor T. We find the constraints on Hubble constant $H_{0}$ and free parameter $n$ with 46 observational Hubble dataset and obtain pretty satisfactory results. The physical features of the model and transitional behavior of equation of state (EOS) parameter are analyzed. We examine the nature of physical parameters and validity of energy conditions as well as stability condition. We also present the Om(z) and statefinder diagnostic analysis for the derived model.
\end{abstract}


\maketitle

\section{Introduction}\label{1}
In 1998, SN Ia observations \cite{Riess/1998} have confirmed that we are lived in an accelerating universe. This discovery has been refined our knowledge and understanding of cosmology and motivates theoretical physicists to think in new direction for modelling the universe. In the literature, numerous models have been investigated to describe the cosmic acceleration by assuming dark energy with repulsive gravity as major content of universe. The cosmological constant $(\Lambda)$ is a prominent candidate of dark energy but it suffers fine tuning and cosmic coincidence problems on theoretical ground so several modifications in general theory of relativity have been introduced in time to time by different researchers such as scalar tensor theory, teleparallel gravity \cite{Einstein/1928}, $f(R)$ and $f(R,T)$ gravity \cite{Harko/2011}. In $f(R,T)$ theory of gravitation, it is assumed that Ricci scalar $(R)$ and trace of energy momentum tensor $(T)$ are coupled with matter Lagrangian \cite{Yadav/2014,Moraes/2017a,Yadav/2018,Yadav/2019,Sharma/2018}. The addition of extra term $T$ in gravitational field will check the possibilities of existence of quantum phenomenons which leads the probabilities of production of particles \cite{Parker/2014}.  Such possibilities may give a clue that there is a bridge between quantum theory of gravity and $f(R,T)$ gravity.Recently Moraes and Sahoo \cite{Moraes/2017a,MoraesSahoo/2017} have constructed a FRW model of hybrid universe governed by non-minimal matter-geometry coupling. In the present paper, we investigate an anisotropic model of singular universe with non-minimal matter geometry coupling in $f(R,T)$ gravity. The field equations have been solved exactly by taking into account the power law variation of scale factor which leads $\Lambda$CDM cosmology. It is worth to mention that $f(R,T)$ theory contributes a significant role to invoke a complete theoretical description for acceleration of universe at late time without aid of exotic matter/energy. Now, there have been a lot of evidences on the various aspects of f(R, T) gravity such as scalar perturbation \cite{Alvarenga/2013}, energy conditions \cite{Alvarenga/2013a,Kiani/2014,Sharif/2013}, thermodynamics \cite{Jamil/2012,Sharif/2012}.\\

The $f(R,T)$ theory of gravitation play a significant role in explaining the current acceleration of universe with resorting the existence of dark energy and dark matter. Now a days, there have been proposed various models on different aspects of $f(R,T)$ gravity and it's functional form \cite{Sharma/2018,Alvarenga/2013,Alvarenga/2013a,Kiani/2014,Jamil/2012,Sharif/2012,Zubair/2016,Sahoo/2017}. In this theory, the interaction of matter with space-time curvature represents a cosmological consequence among different matter components \cite{Sahoo/2017a}. However the role of violation of energy-momentum conservation (EMC) in this theory have not yet been studied properly but one significant contribution of violation of EMC leads to accelerated expansion in the cosmological models of $f(R,T)$ gravity \cite{Josset/2017}. Some pioneer studies of stellar objects within the framework of $f(R,T)$ theory of gravity are given in references \cite{Yousaf/2016,Yousaf/2018,Yousaf/2016a,Yousaf/2019,Das/2016, Maurya/2017}. Motivated by the above discussion, here we confine ourselves to investigate a cosmological model of non-minimal matter-geometry coupling in Bianchi V space-time in $f(R,T)= f_{1}(R)+f_{2}(R)f_{3}(T)$ gravity. Bianchi V universe is a natural generalization of isotropic and flat model of universe, is of particular
interest because it describes the homogeneous and an-isotropic universe with different scale factors in each spatial axes. Despite of recent elaboration, in the literature, several authors have
constructed different cosmological model in different physical contexts \cite{Yadav/2014,Kumar/2011,Saha/2004,Yadav/2016,Yadav/2012} in Bianchi V space-time. Recently, Mishra et al. \cite{Mishra/2018} have investigated string cosmological model with dark energy in anisotropic Bianchi V space-time. Some useful applications of anisotropy features of present universe are given in refernces \cite{Mishra/2015,Mishra/2015a,Mishra/2018a,Tripathy/2017,Koivisto/2006}.\\

In the present paper, we confine ourselves to study coupling of matter and geometry power law in Bianchi-V space-time. Other interesting reference that show the wide applicability of Bianchi-V model of universe are \cite{saha/2006,yadav/2011,yadav/2012a,kumar/2011a}. The paper is organized as follow: in section \ref{2}, we present the basic of $f(R,T)$ theory of gravitation and scale factors for Bianchi V space-time and compute the cosmological parameters. In section \ref{3} we confront the theoretical predictions of our model with observational Hubble data. Section \ref{4} represents the physical behavior of derived model. Section \ref{5} deals with the violation/validation of energy conditions and stability condition of the derived model. The Om(z) and state-finder diagnosis for Bianchi-V universe within the framework of $f(R,T)$ gravity is done in section \ref{6} \& \ref{7}. Finally in section \ref{8}, we have summarized our results.

\section{The basic of $f(R,T)$ theory of gravitation and scale factors}\label{2}
Bianchi-V space-time is read as
\begin{equation}
\label{ref1}
 ds^{2}=-c^{2}dt^{2}+A(t)^{2}dx^{2}+e^{2\alpha x}\left(B^{2}dy^{2}+C^{2}dz^{2}\right)
\end{equation}
Here, $A(t)$, $B(t)$, $C(t)$ are scale factor along $x, y$ and $z$ axes and $\alpha$ is a constant.\\
The action in $f(R,T)$ gravity is given by
\begin{equation}
 \label{basic}
S = \frac{1}{16\pi}\int{d^{4}x\sqrt{-g}f(R,T)} +\int{d^{4}x\sqrt{-g}L_{m}}
\end{equation}
where $g$ and $L_{m}$ are the metric determinant and matter Lagrangian density respectively.\\
The gravitational field of $f(R,T)$ gravity is given by
\[
\left[f_{1}^{\prime}(R)+f_{2}^{\prime}(R)f_{3}^{\prime}(T)\right]R_{ij}-\frac{1}{2}f_{1}^{\prime}(R)g_{ij}+
\]
\[
(g_{ij}\nabla^{i}\nabla_{j}-\nabla_{i}\nabla_{j}) \left[f_{1}^{\prime}(R)+f_{2}^{\prime}(R)f_{3}^{\prime}T\right]=
\]
\begin{equation}
\label{basic1}
[8\pi+f_{2}^{\prime}(R)f_{3}^{\prime}(T)]T_{ij}+f_{2}(R)\left[f_{3}^{\prime}(T)p+\frac{1}{2}f_{3}(T)\right]g_{ij}
\end{equation}
Here, $f(R,T) = f_{1}(R)+f_{2}(R)f_{3}(T)$ and primes denote derivatives with respect to the arrangement.\\
Following Moraes and Sahoo \cite{Moraes/2017a}, we assume $f_{1}(R) = f_{2}(R) = R$ and $f_{3}(T) = \zeta T$, with
$\zeta$ as a constant. Thus the equation (\ref{basic1}) yields
\begin{equation}
 \label{basic2}
G_{ij} = 8\pi T^{(eff)}_{ij} = 8\pi(T_{ij}+T^{(ME)}_{ij})
\end{equation}
where, $T^{(eff)}_{ij}$, $T_{ij}$ and $T^{(ME)}_{ij}$ represent the effective energy momentum tensor, matter energy momentum tensor and extra energy term due to trace of energy-momentum tensor respectively. The extra energy term is read as
\begin{equation}
 \label{basic3}
T^{(ME)}_{ij}=\frac{\zeta R}{8\pi}\left(T_{ij}+\frac{3\rho-7p}{2}g_{ij}\right)
\end{equation}
By applying the Bianchi identities in equation (\ref{basic2}) yields
\begin{equation}
 \label{basic4}
\nabla^{i}T_{ij} = -\frac{\zeta R}{8\pi}\left[\nabla^{i}(T_{ij}+pg_{ij})+
\frac{1}{2}g_{ij}\nabla^{i}(\rho-3p)\right]
\end{equation}
For the line element (\ref{ref1}), the field equation (\ref{basic2}) can be written as
\begin{equation}
\label{fe1}
 \frac{\ddot{B}}{B}+\frac{\ddot{C}}{C}+\frac{\dot{B}\dot{C}}{BC}-\frac{\alpha^{2}}{A^{2}} = -8\pi p^{(eff)}
\end{equation}
\begin{equation}
\label{fe2}
 \frac{\ddot{C}}{C}+\frac{\ddot{A}}{A}+\frac{\dot{A}\dot{C}}{AC}-\frac{\alpha^{2}}{A^{2}} = -8\pi p^{(eff)}
\end{equation}
\begin{equation}
\label{fe3}
 \frac{\ddot{A}}{A}+\frac{\ddot{B}}{B}+\frac{\dot{A}\dot{B}}{AB}-\frac{\alpha^{2}}{A^{2}} = -8\pi p^{(eff)}
\end{equation}
\begin{equation}
\label{fe4}
 \frac{\dot{A}\dot{B}}{AB}+\frac{\dot{B}\dot{C}}{BC}+\frac{\dot{C}\dot{A}}{CA}+\frac{\alpha^{2}}{A^{2}} = 8\pi \rho^{(eff)}
\end{equation}
Here, $\rho^{(eff)} = \rho + \rho^{(ME)} = \rho -\frac{3\zeta}{8\pi}\left(\frac{\ddot{a}}{a}+
\frac{\dot{a}^{2}}{a^{2}}\right)(3\rho - 7p)$,
$p^{(eff)} = p + p^{(ME)} = p +\frac{9\zeta}{8\pi}\left(\frac{\ddot{a}}{a}+
\frac{\dot{a}^{2}}{a^{2}}\right)(\rho - 3p)$ and $a= (ABC)^{\frac{1}{3}}$ is average scale factor.\\

The above equations (\ref{fe1})-(\ref{fe4}) can also be written as
\begin{equation}
\label{ge}
\frac{(ABC)^{\ddot{}}}{ABC} = 12\pi(\rho^{(eff)}- p^{(eff)})
\end{equation}
We define Hubble's
parameter $(H)$ in connection with average scale factor as follow:
\begin{equation}
\label{H}
 H=\frac{\dot{a}}{a}
\end{equation}
Following, Sharma et al \cite{Sharma/2018}, we have
\begin{equation}\label{a(t)}
 a = (nDt)^{\frac{1}{n}}
\end{equation}
where $n$ \& $D$ are non zero positive constants.\\
Solving equations (\ref{fe1}), (\ref{fe2}), (\ref{fe3}) and (\ref{dp}), we get
\begin{equation}\label{x}
 A(t) = (nDt)^{\frac{1}{n}}
\end{equation}
\begin{equation}\label{y}
 B(t) = \xi(nDt)^{\frac{1}{n}}exp\left[\frac{\delta}{D(n-3)}(nDt)^{\frac{n-3}{n}}\right]
\end{equation}
\begin{equation}\label{z}
 C(t) = \xi^{-1}(nDt)^{\frac{1}{n}}exp\left[\frac{-\delta}{D(n-3)}(nDt)^{\frac{n-3}{n}}\right]
\end{equation}
Here $\xi$ and $\delta$ are constants.\\
\section{Confrontation with Hubble data}\label{3}
The scale factor $(a)$ and the redshift (z) are connected through the following relation 
\begin{equation}
\label{a(z)}
a = \frac{a_{0}}{1+z}
\end{equation} 
where $a_{0}$ is the present value of scale factor.\\
Thus, from eqs (\ref{H}), (\ref{a(t)}) and (\ref{a(z)}), the Hubble's parameter in term of redshift may be expressed as
\begin{equation}
\label{H(z)}
H(z) = H_{0}(1+z)^{n}
\end{equation}
where $H_{0}$ is the present value of Hubble's parameter.\\

The differential age of the galaxies are used for the observational Hubble data (OHD) by following relation\cite{Riess/2011,Akarsu/2014}
\begin{figure*}[thbp]
\begin{center}
\begin{tabular}{rl}
\includegraphics[width=7.5cm]{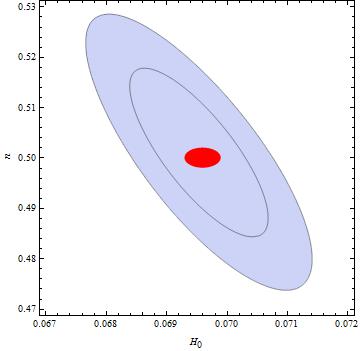}
\end{tabular}
\end{center}
\caption{The likelihood contours at $1\sigma$ and $2 \sigma$ confidence levels around the best fit values points (0.0695, 0.501) in $H_{0} - n$ plane. The present value of $H_{0}$ is estimated as $0.0695~Gyrs^{-1} \sim 67.98~km s^{-1}Mpc^{-1}$.} 
\end{figure*}
\begin{figure*}[thbp]
\begin{center}
\begin{tabular}{rl}
\includegraphics[width=7.5cm]{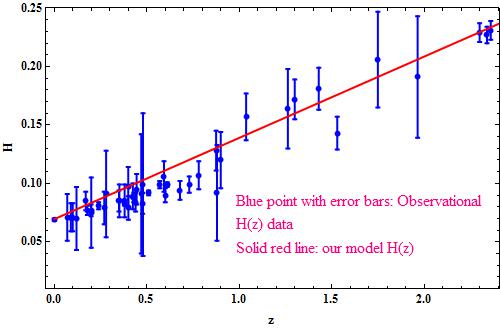}
\end{tabular}
\end{center}
\caption{Fitting of our model with 46 OHD points}.
\end{figure*}
\begin{equation}
\label{H(z)-2}
H(z) = -\frac{1}{1+z}\frac{dz}{dt}
\end{equation} 
We consider the latest OHD data points in the redshift range $0 \leq z \leq 2.36$ with the corresponding standard deviation $\sigma_{H}$. The OHD data points are compiled in references \cite{Akarsu/2014,Zhang/2012,Yadav/2019a}.\\

The mean value of model parameters $H_{0}$ and $n$ are determined by minimizing 
\begin{equation}
\label{chi}
\chi^{2}_{OHD} = \sum_{i=1}^{46}\frac{[H_{th}(z_{i})-H_{obs}(z_{i})]^2}{\sigma_{H}^{2}(z_{i})}
\end{equation}
where $H_{th}$ denotes the model based theoretical value of Hubble's parameter, $H_{obs}$ stands for observed values of Hubble's parameter and $\sigma_{H}$ is the standard errors in each observed value. The likelihood contours in $H_{0} - n$ plane at $1\sigma$ and $2\sigma$ confidence levels of our model is shown in Figure 1. The best fit values of $n$ and $H_{0}$ are obtained as $n = 0.501$  and $H_{0} = 0.0695~Gyrs^{-1}~ \sim~67.98~km s^{-1} Mpc^{-1} $ with minimum $\chi^{2} = 1.798$. Figure 2 shows Fitting of our model with 46 OHD points . The figure clearly shows a pretty of model under consideration with OHD points for n = 0.501  and $H_{0} = 67.98$ with in the range $0 \leq z \leq 2.36$. Thus we set $n = 0.501$ for graphical analysis of all physical parameters of derived model.\\
\section{Physical behavior of the Model}\label{4}
Substituting equations (\ref{x}) - (\ref{z}) in equations (\ref{fe3}) and (\ref{fe4}), the expressions for
$p^{(eff)}$ and $\rho^{(eff)}$ are respectively given by
\begin{equation}
 \label{pe}
8\pi p^{(eff)} = D^{2}(2n-3)(nDt)^{-2}-\xi^{2}(nDt)^{\frac{-6}{n}}+\alpha^{2}(nDt)^{-\frac{2}{n}}
\end{equation}
\begin{equation}
 \label{rhoe}
8\pi\rho^{(eff)} = 3D^{2}(nDt)^{-2}-\xi^{2}(nDt)^{\frac{-6}{n}}-3\alpha^{2}(nDt)^{-\frac{2}{n}}
\end{equation}
\begin{figure*}[thbp]
\begin{tabular}{rl}
\includegraphics[width=6cm]{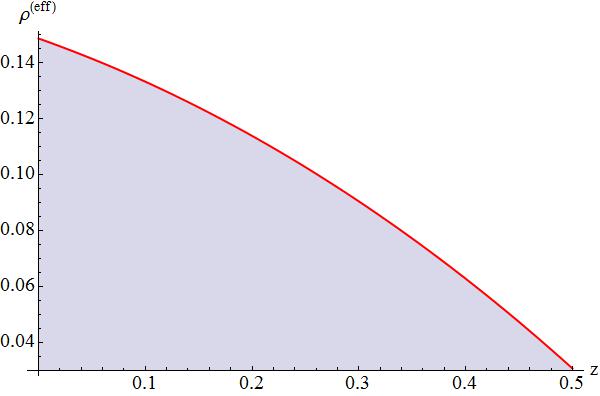}
\includegraphics[width=6cm]{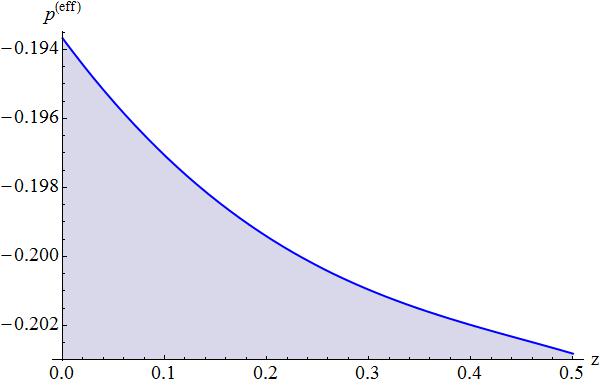}\\
\end{tabular}
\caption{Effective energy density and effective pressure vs. z with $n = 0.501$, $\alpha = 0.35$, $D = 0.65$ and $\xi = 0.05$.}
\end{figure*}

The expressions for volume scalar $(V)$ is read as
\begin{equation}
\label{V}
V = (nDt)^{\frac{3}{n}}
\end{equation}
From equation (\ref{V}), it is evident that at t = 0, the spatial volume vanishes and at $t \rightarrow \infty$. Fig. 4 shows the expanding behavior of volume scalar V with passage of time t.\\
\begin{figure*}[thbp]
\begin{tabular}{rl}
\includegraphics[width=10cm]{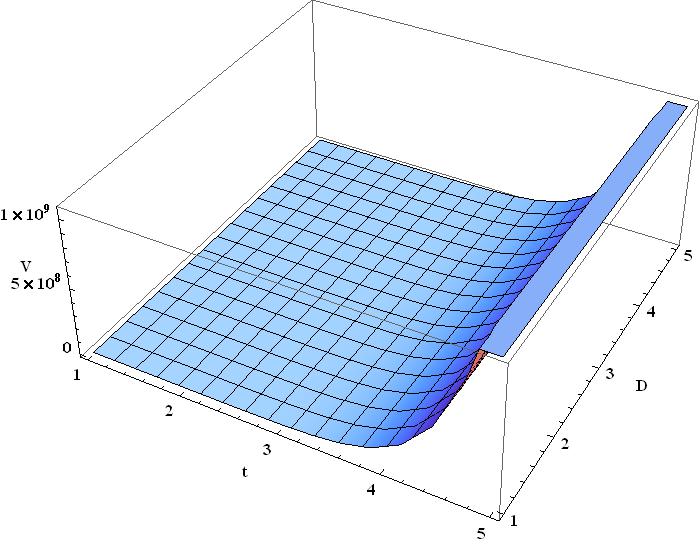}
\end{tabular}
\caption{Spatial volume vs. t with $n = 0.42$}.
\end{figure*}

The expansion scalar $(\theta)$ and deceleration parameter are obtained as
\begin{equation}
\label{expansion}
\theta = 3D(nDt)^{-1}
\end{equation}
\begin{equation}
\label{dp}
q = n-1
\end{equation}
The anisotropic parameter is read as
\[
\Delta = \frac{1}{9H^{2}}\left[\left(\frac{\dot{A}}{A}-\frac{\dot{B}}{B}\right)^{2}+\left(\frac{\dot{B}}{B}-\frac{\dot{C}}{C}\right)^{2}+\left(\frac{\dot{C}}{C}-\frac{\dot{A}}{A}\right)^{2}\right]
\]
\begin{equation}
\label{anisotropic}
 = \frac{2\delta^{2}}{3D^{2}}(nDt)^{-\frac{2(3-n)}{n}}\;\;\;\;\;\;\;\;\;\;\;\;\;\;\;\;\;\;\;\;\;\;\;\;\;\;\;\;\;\;\;\;\;\;\;\;\;\;\;\;\;\;\;\;\;\;\;
\end{equation} 
Fig. 3 depicts the behaviour of effective energy density $\rho^{(eff)}$ and effective pressure $p^{(eff)}$ versus red-shift z for $n =0.501$. From equation (\ref{dp}), it is evident that for $0\leq n \leq 1$, the deceleration parameter evolves in the range $-1 \leq q \leq 0$. For $n = 0.501$ in equation (\ref{dp}), we obtain $q = -0.499$, which is close to the observed value of deceleration parameter at present epoch \cite{cunha/2009}. Thus we have constrained $n = 0.501$, in the graphical analysis of the physical parameters of derived model. The behaviour of anisotropic parameter versus time is depicted in Figure 5 which clearly shows that anisotropic parameter starts with high value but diminishes at present i.e. anisotropic universe becomes isotropic at larger time.\\
\begin{figure*}[thbp]
\begin{center}
\begin{tabular}{rl}
\includegraphics[width=7.5cm]{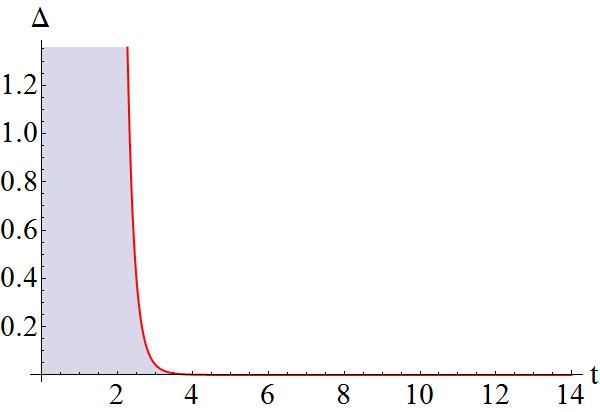}
\end{tabular}
\end{center}
\caption{The behaviour of anisotropic parameter with respect to time for $n = 0.501$ and $D = 0.65$.}
\end{figure*}

It is important to note that for $n \neq 0$, the spatial volume of derived universe is zero and expansion scalar is infinite at initial epoch $t = 0$ which shows that the universe starts expanding with infinite rate at $t = 0$. On later times, we also observe that $\frac{dH}{dt} = 0$ and $q = -1$ which implies that the derived universe has greatest value of Huble's parameter and fastest rate of expansion at $t = \infty$. \\

The expressions for $p^{(eff)}$ and $\rho^{(eff)}$ in terms of red-shift (z) are computed as following
\begin{equation}
\label{pz}
8\pi p^{(eff)} = (2n-3)D^{2}(1+z)^{2n}-\xi^{2}(1+z)^{6}+\alpha^{2}(1+z)^{2}
\end{equation} 
\begin{equation}
\label{rhoz}
8\pi\rho^{(eff)} = 3D^{2}(1+z)^{2n}-\xi^{2}(1+z)^{6}-3\alpha^{2}(1+z)^{2}
\end{equation}
The equation of state parameter (EOS) parameter $\omega^{(eff)}$ can be obtained from equations (\ref{pz}) and (\ref{rhoz}) as
\begin{equation}
\label{omega}
\omega^{(eff)} = -1+\frac{2nD^{2}(1+z)^{2n}+2\xi^{2}(1+z)^{6}+2\alpha^{2}(1+z)^{2}}{3D^{2}(1+z)^{2n}+\xi^{2}(1+z)^{6}-\alpha^{2}(1+z)^{2}}
\end{equation}
\begin{figure*}[thbp]
\begin{tabular}{rl}
\includegraphics[width=8cm]{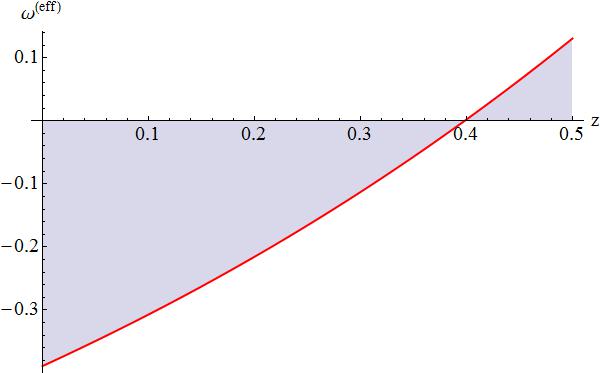}
\end{tabular}
\caption{$\omega^{(eff)}$ vs. z with $n = 0.501$, $\alpha = 0.35$, $D = 0.65$ and $\xi = 0.05$.}
\end{figure*}
\begin{figure*}[thbp]
\begin{tabular}{rl}
\includegraphics[width=10cm]{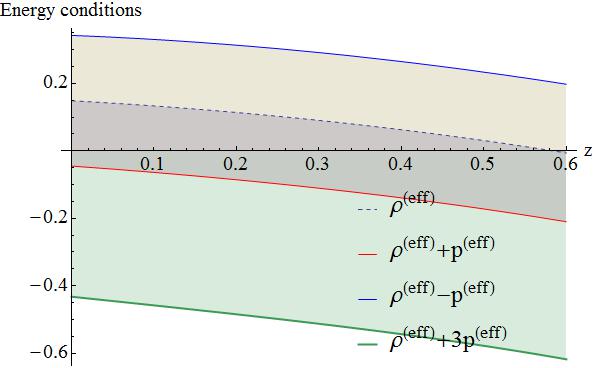}
\end{tabular}
\caption{Single plot of Energy conditions vs. z with $n = 0.501$, $\alpha = 0.35$, $D = 0.65$ and $\xi = 0.05$.}
\end{figure*}
\begin{figure*}[thbp]
\begin{tabular}{rl}
\includegraphics[width=6cm]{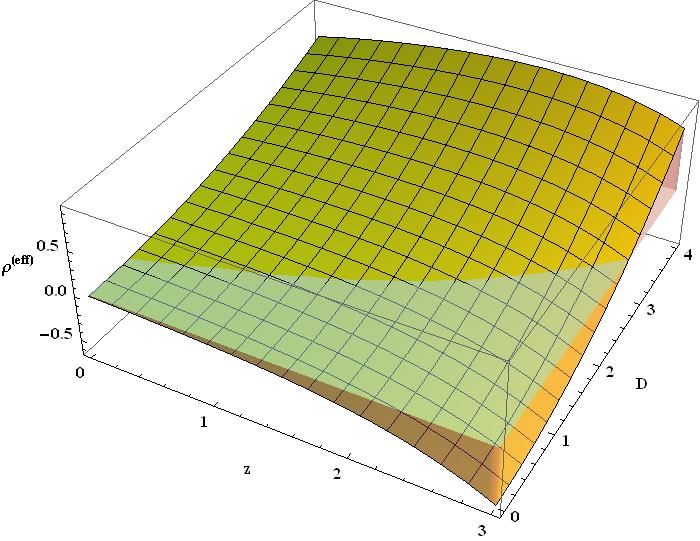}
\includegraphics[width=6cm]{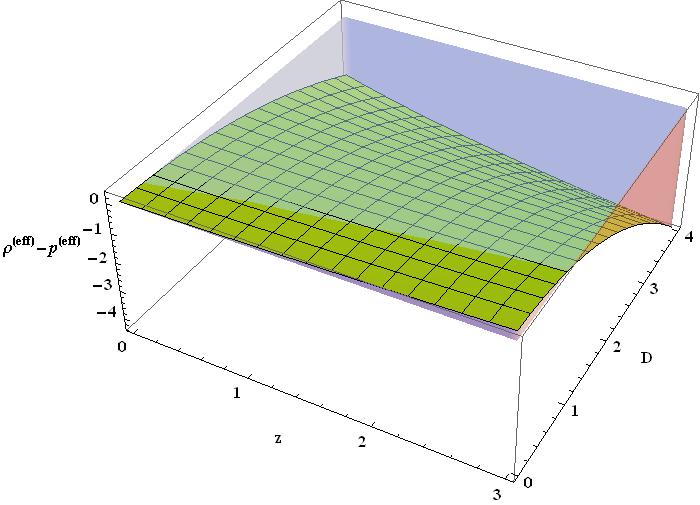}\\
\includegraphics[width=6cm]{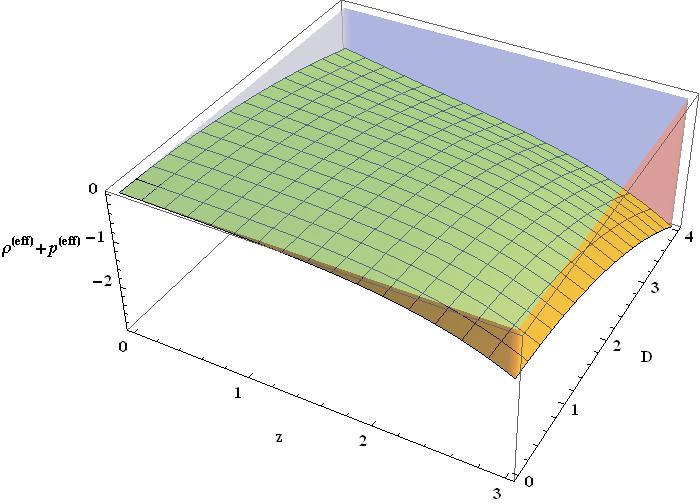}
\includegraphics[width=6cm]{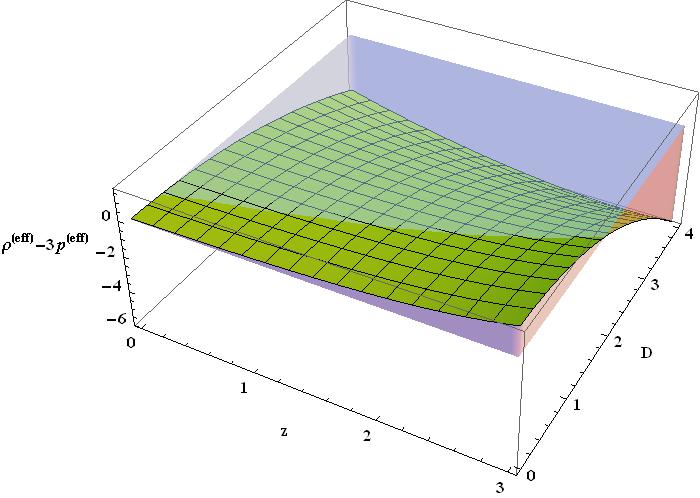}\\
\end{tabular}
\caption{Energy conditions for $0\leq z \leq 3$ and $0\leq D \leq 4$ .}
\end{figure*}
In Fig. 6, we have graphed the evolution of EOS parameter as a function of redshift z. The EOS parameter exhibits transitional behavior from positive to negative at present epoch. This transitional behavior $\omega^{(eff)}$ fits with recent data obtained by SNL3 obervation\cite{Feng/2005}. Some other reconstruction of the EOS parameter from different observational data sets in particular the high redshift Lyman-$\alpha $ forest mesurement favors a non constant dynamical dark energy. From Figure 2, it is clearly visible that $\omega^{(eff)}$ of derived model evolves from radiation dominated phase $(\omega^{(eff)} = \frac{1}{3})$ to a matter dominated phase $(\omega^{(eff)} = 0)$ and then to a dark energy driven accelerated phase $(\omega^{(eff}) < -\frac{1}{3})$ \cite{Sahoo/2018}. At present epoch (z = 0), our model predicts that the current universe is in accelerating phase.\\        
\section{Energy condition and stability}\label{5}
The energy conditions are important ingredient to interpret many aspects of the universe including expansion and singularity postulates. The authors of references (Riess et al.\cite{Riess/1998} , Perlmutter et al. \cite{Perlmutter/1999}) had shown that energy conditions are required to study and strong energy condition (SEC) must be abandoned \cite{Barcelo/2002}. The violation of SEC is generally due to the anti gravity matter such as dark energy present in the universe. For normal matter, all the energy conditions including SEC must be validate. In the derived model, Fig. 7 shows the validation and violation of energy conditions for particular choice of free parameters and satisfies the following.\\

$\rho^{(eff)}\geq 0$, $\rho^{(eff)} - p^{(eff)} \geq 0$, $\rho^{(eff)} + p^{(eff)} \leq 0$ and $\rho^{(eff)} - 3p^{(eff)} \leq 0$\\

Thus the derived model validates the weak energy condition (WEC) and dominant energy condition (DEC) and violates the Null energy condition (NEC) and strong energy condition (SEC) with specific choice of the free parameters and explore the possibilities of acceleration of universe with $f(R,T)$ gravity formalism. In $f(R,T)$ theory of gravity the consequences of energy conditions have been analyzed by Sharif et al. \cite{Sharif/2013a}.~ Fig. 8 exhibits the detail analysis of validation of energy conditions within the range $0\leq z \leq 3$ and $0\leq D \leq 4$. From Fig. 8, it is evident that WEC and NEC are satisfied within the range $0\leq z \leq 1$ and $0\leq D \leq 1$ while DEC and SEC are violated. In the range $1\leq z \leq 1$ and $1\leq D \leq 2$, only WEC is validated but NEC, DEC as well as SEC are violated. None of energy conditions are satisfied for $z\geq 1$ and $D\geq 2$ which may impose the restriction on the choice of free parameters.    
\begin{figure*}
\includegraphics[width=7.5cm]{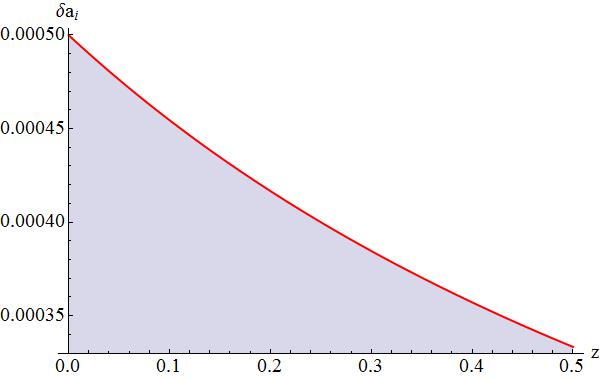}
\caption{Validation of stability condition for $n=0.501$.}\label{fig9}
\end{figure*}
One may check the stability of the derived solution with respect to the perturbation of the space-time \cite{Saha/2012}. For this purpose, we have assumed the perturbations of volume scalar, directional Hubble factors and mean Hubble factor as
\begin{equation}
\label{stability2}
V\rightarrow V_{B}+V_{B}\sum_{i}\delta b_{i}, \,\,\,\,\,\theta_{i} \rightarrow \theta_{Bi}+\sum_{i}\delta b_{i}, \,\,\,\,
\theta \rightarrow \theta_{B}+\frac{1}{3}\sum_{i}\delta b_{i} 
\end{equation}
where $\delta b_{i}$, $V_{B}$ denote the perturbation term and background spatial volume.\\

For derived model, the perturbation term is read as
\begin{equation}
\label{st8}
\delta b_{i} = \zeta_{2}-\zeta_{1}exp(-3/n)nt
\end{equation}
where $\zeta_{1}$ and $\zeta_{2}$ are the constants of integration.\\
Thus, the actual fluctuations $\delta a_{i} = a_{Bi}\delta b_{i}$ is computed as
\begin{equation}
\label{st9}
\delta a_{i} = \zeta_{2}(nDt)^{\frac{1}{n}}-\zeta_{1}D^{\frac{1}{n}}exp(-3/n)(nt)^{\frac{n+1}{n}}
\end{equation}
In terms of red-shift, the actual fluctuation is computed as
\begin{equation}
\label{st10}
\delta a_{i} = \frac{\zeta_{2}}{1+z}-\zeta_{1}D^{1/n}exp(-3/n)D(1+z)^{\left(-\frac{n+1}{n}\right)}
\end{equation} 

Fig. 9 depicts the behaviour of actual fluctuations with respect to red-shift. In past the actual fluctuation is null but it increases minimally on later time.\\ 
\section{Om diagnostic analysis}\label{6}
The Om(z) parameter \cite{Sahooetal/2018} is given by
\begin{equation}
\label{om1}
Om(z) = \frac{\left[\frac{H(z)}{H_{0}}\right]^{2}-1}{(1+z)^{3}-1}
\end{equation}

The Om(z) parameter of derived model is given by
\begin{equation}
\label{om2}
Om(z) = \frac{(1+z)^{-2n}-n^{4}H_{0}^{2}D^{2}}{n^{4}H_{0}^{2}D^{2}\left[(1+z)^{3}-1\right]}
\end{equation}

In the literature, Om(z) diagnostic analysis is useful to modelling the dynamics of dark energy \cite{Sahni/2008}. In comparision with the state finder diagnosis, the Om parameter involves only first derivative of scale factor. The positive, negative and zero values of Om(z) parameter consistent with phantom, quintessence and $\Lambda$CDM dark energy models respectively \cite{Sahooetal/2018,Shahalam/2015,Sahoo/2018}. Fig. 10 depicts the behavior of Om(z) parameter against z of derived model. The left panel of Fig. 10 shows the dynamics of Om(z) parameter for particular value of D = 0.65 and n = 0.501 whereas the right panel explores the nature of Om(z) parameter in the range $0\leq D \leq 2$. In both the panel, Om(z) parameter is negative and monotonically increasing within the interval $0\leq z \leq 3$ which also suggests that at present the universe is in accelerating mode.\\
\begin{figure*}[thbp]
\begin{tabular}{rl}
\includegraphics[width=6cm]{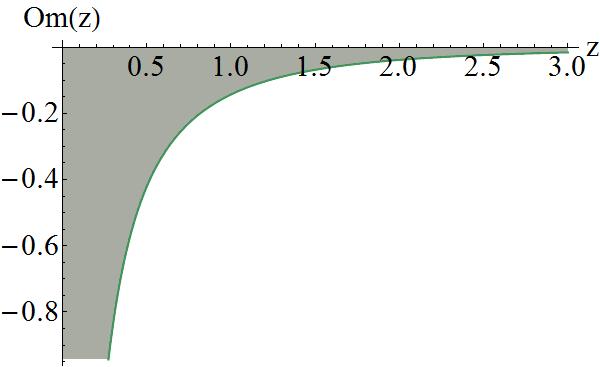}
\includegraphics[width=6cm]{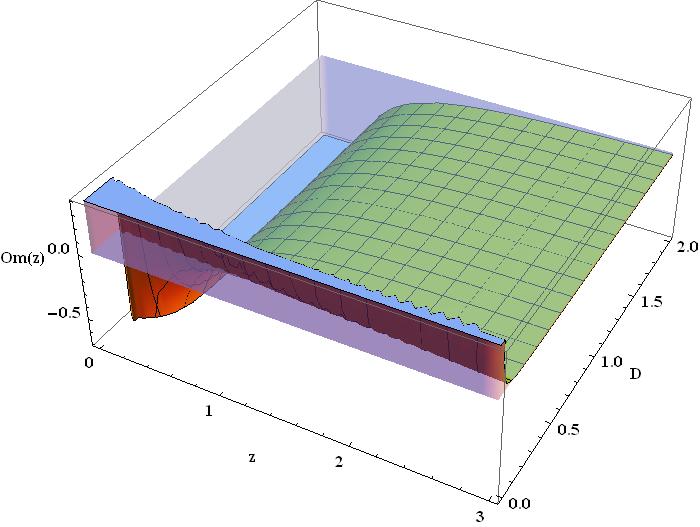}\\
\end{tabular}
\caption{Om(z) versus z with $H_{0} = 67.98~km s^{-1}Mpc^{-1}$.}
\end{figure*}
\section{Statefinder diagnostic}\label{7} 
In 2003, Sahni et al. \cite{Sahni/2003} have introduced a pair of parameters (r,s) called statefinder which represents the geometrical diagnostic of dark energy. The statefinders are defined as
\begin{equation}
\label{rs-1}
r=\frac{\dot{\ddot{a}}}{aH^3}
\end{equation}
\begin{equation}
\label{rs-2}
s=\frac{r-1}{3(-\frac{1}{2}+q)}
\end{equation}
Here, one can not choose $q\neq\frac{1}{2}$.\\
\begin{figure*}[thbp]
\begin{center}
\begin{tabular}{rl}
\includegraphics[width=7.5cm]{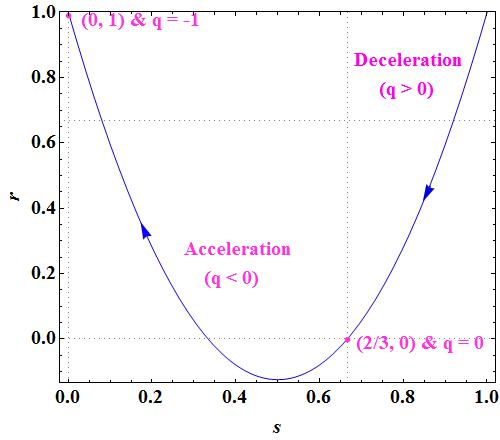}
\end{tabular}
\end{center}
\caption{Dynamics of s - r plane}
\end{figure*}

From equations (\ref{a(t)}), (\ref{dp}), (\ref{rs-1}) and (\ref{rs-2}), we have
$r = 2q^2 + q$ and $s = \frac{2}{3}(q+1)$\\
The trajectories of s -r plane  is shown in Fig. 11. From Fig. 11, we notice that at (s,r) = (1,1), $q > 0$ represents the deceleration and (s,r) = (0,1), $q = -1$ represents the location of flat $\Lambda$CDM. A transition from decelerating mode to accelerating mode occurs at (s,r) = (2/3,0). Thus the trajectories in s - r plane of derived model starts from standard cold dark matter model and finally approaches $\Lambda$CDM model. 
Thus the statefinder diagnostic of derived model is not new but it represents similar result as obtained in previous investigations \cite{Kumar/2012}. 
\section{Conclusion}\label{8}
In this paper, we have studied the coupling of non-minimal matter-geometry in Bianchi-V space time. The derived model is based on $f(R,T)$ gravity and it's functional form $f_1(R)=f_2(R)=R$ and $f_3(T)=\zeta T$ with power law. The main features of the paper is as follows:\\
\begin{itemize}
\item~ In the present model, the effective energy density remains positive. We have analyzed the validity and violation of energy conditions both analytically as well as graphically. In the derived model, WEC and DEC are satisfied whereas NEC and SEC are violated for particular choice of free parameters. The validations of NEC and WEC show that derived model is physically reasonable and coexist with $f(R,T)$ solution whose energy-momentum tensor is not conserved. The violation of SEC in an accelerated universe is general which is driven by an exotic matter in general theory of relativity. Indeed the violation of DEC and SEC favor the possibilities of acceleration without aid of exotic matter in $f(R,T)$ theory \cite{Sharma/2018, MoraesSahoo/2017, Barcelo/2002}. 
\item~ At $t = 0$, the scale factors vanish and expansion scalar becomes infinity. Thus the derived model has singularity at $t = 0$ and it is point type singularity.
\item~ The constraints on $H_{0}$ and $n$ clearly indicates that the derived model fits best to the OHD points and the estimate of Hubble's constant is in close agreement with Riess et al \cite{Riess/2011}. 
\item~ The derived model describes the dynamics of universe from big bang to present epoch in the modified theory of gravity and explore the possibilities of accelerating universe with non-exotic fluid as the source of matter.
\item~ The present age of the universe is computed as
$$T_{0} = \frac{1}{n}H_{0}^{-1}$$
For $0 \leq n \leq 1$, the present age of universe will increase.
\item~ The EOS parameter of derived model shows the transitional behavior and it evolves from a radiation dominated phase to a matter dominated phase and then to a dark energy driven accelerated phase.
\item~ The Om(z) parameter of derived model is negative which is consistent with accelerating universe.     
\item~ The statefinder analysis shows that derived model approaches the dynamics of $\Lambda$CDM model in future.
\end{itemize} 

In summary, the present model describes the possibility that acceleration of universe may be geometrical contribution of the matter inside it and $f(R,T)= R + \zeta R T$ gravity model as alternative to cosmic acceleration - one of the significant feature of $f(R,T)$ gravity. The directional scale factors and spatial volume of the derived model vanish at $t = 0$. As $t \rightarrow \infty$, the scale factors diverse while effective energy density and effective pressure approach zero value. Thus all matter and radiation is concentrated at
the big bang singularity and the cosmic expansion is driven by the big bang impulse i.e. the derived solution represents the model of singular universe which initially starts from big-bang at $t = 0$ and describes the acceleration of universe at present epoch.\\     


\end{document}